\documentclass[a4paper,11pt]{article}

\pdfoutput=1 
\usepackage{cite}
\usepackage{jheppubmod}
\usepackage{enumitem}
\usepackage{amsmath}
\usepackage{physics}
\usepackage{graphicx}
\usepackage{amsfonts} 
\usepackage{amssymb}
\usepackage[dvipsnames, svgnames, x11names ]{xcolor}
\usepackage{color}
\usepackage{braket}
\graphicspath{ {./1 /} }
\usepackage{subcaption}
\usepackage{graphicx}
\usepackage[T1]{fontenc} 
\usepackage{tikz}
\usepackage{physics}
\usepackage{amsmath}
\usepackage{dsfont}
\usepackage{graphicx}
\usepackage{bbold}
\graphicspath{ {./1 /} }

\title{Algebra of Operators in an AdS-Rindler Wedge}

\author{Eyoab Bahiru}
 \affiliation{SISSA, International School for Advanced Studies, Via Bonomea 265, 34136 Trieste, Italy}
 \affiliation{INFN, National Institute for Nuclear Physics,  Sezione di Trieste,
Via Valerio 2, 34127 Trieste, Italy}
 \affiliation{ICTP, International Centre for Theoretical Physics, Via
Strada Costiera 11, 34151, Trieste, Italy}
\emailAdd{ebahiru@sissa.it}

\date{\today}

\abstract{
We discuss the algebra of operators in AdS-Rindler wedge, particularly in AdS$_{5}$/CFT$_{4}$. We explicitly construct the algebra at $N=\infty$ limit and discuss its Type III$_{1}$ nature. We will consider $1/N$ corrections to the theory and using a novel way of renormalizing the area of Ryu-Takayanagi surface, describe how several divergences can be renormalized and the algebra becomes Type II$_{\infty}$. This will make it possible to associate a density matrix to any state in the Hilbert space and thus a von Neumann entropy.  
}

\begin{document}
\maketitle
\section{Introduction}

There has been a recent interest in studying algebra of operators in perturbative limits of quantum gravity in the context of AdS/CFT. Leutheusser and Liu studied the $N=\infty$ limit of the thermofield double state above Hawking-Page temperature \cite{Leutheusser:2021frk,Leutheusser:2021qhd}. They noted that the algebra of operators exterior to the black hole horizon form what is called Type III$_{1}$ von Neumann algebra. From the boundary perspective, the algebra elements are the single trace operators whose thermal one point function is subtracted,
$$ \Tilde{O}=O-\langle O \rangle .$$
These operators are generalized free fields with $O(1)$ two point functions. Using properties of Type III$_{1}$ algebras, they were able to propose a natural time like evolution operator for an observer falling into the black hole.

Witten later proposed to include the Hamiltonian into the algebra \cite{Witten:2021unn}. In the strict $N\rightarrow\infty$ limit, 
$$ [U,\Tilde{O}]=0$$
where $U$ is the properly normalized Hamiltonian with a definite large $N$ limit, $U=\frac{H-\langle H \rangle}{N}$. Thus the algebra at this level is in fact a tensor product of the GFF algebra and the algebra of bounded functions of $U$, $\mathcal{A}_{\tilde{O}}\otimes \mathcal{A}_{U}$ and still a Type III$_{1}$ algebra.

Another property of Type III$_{1}$ von Neumann algebras is that one can extend it to include the outer automorphism of the algebra and changes it to a Type II$_{\infty}$ von Neumann algebra. This algebra will be acting on a different Hilbert space and does not describe the original system\footnote{Still, by doing an additional crossed product by the dual group of the automorphism to this new algebra, one can get to the tensor product of the original Type III$_{1}$ algebra and an algebra of bounded functions on $\mathcal{L}^{2}(G)$. To study Type III$_{1}$ algebras like this was the original purpose of the crossed product construction.}. But, as one includes $1/N$ corrections to the operators exterior to the black hole, it was found that the center mode $U$ will generate an outer automorphism for the algebra $\mathcal{A}_{\tilde{O}}$. Thus the new Type II$_{\infty}$ algebra in fact describes the algebra of operators as one backs away form strict $N\rightarrow\infty$ limit and considers $1/N$ corrections. Entropy can be associated to this algebra of operators and it can  be interpreted as the generalized entropy for the black hole.
This construction is also done for maximally extended Schwarzschild black hole in flat space-time and deSitter space-time \cite{Chandrasekaran:2022cip}.

The purpose of this paper is to analyze this construction in the case of AdS-Rindler wedge. This corresponds to a spherical region in the boundary CFT and one needs to analyze single trace operators in this sub-region. They form a Type III$_{1}$ algebra and there is a center for these operators. But there is a difference with the previous discussion in that the two point function of the center mode will still be divergent after one naively normalizes it, in a similar way $U$ is defined. In fact the two point function is given by the area of the horizon. We will renormalize this area and thus the two point function then we will be able to proceed with the construction that goes parallel to the previous discussions for deSitter and black holes in AdS and flat space-times. This will also provide a new renormalization scheme for the area Ryu-Takayanagi surface of a spherical ball subregion of boundary CFT$_{d}$ in $d \geq 2$. 

\section{The Type III$_{1}$ nature of the AdS-Rindler wedge}

Consider two time-like separated points in the boundary CFT, $p$ and $q$, where $p$ is to the future of $q$. The intersection of the causal past of $p$ and the causal future of $q$ is defined to  be the causal wedge $D^{p}_{q}$. To describe the physics in this diamond, it is enough to know the physical data on a space-like slice in the diamond called the Cauchy surface \footnote{This is not true in general, for instance in the strict large $N$ limit of the boundary CFT.}. This Cauchy slice can be chosen to be finite spherical region at some time slice in the boundary. The causal wedge of $D^{p}_{q}$, what is called the AdS-Rindler wedge, is the bulk region that is causally connected $D^{p}_{q}$. In other words, the fields in this bulk region can be explicitly expressed by evolving the boundary operators in $D^{p}_{q}$ in to the bulk using the bulk equations of motion at the time slice\cite{Hamilton:2005ju,Hamilton:2006az,Almheiri:2014lwa}. Since one can think of the causal wedge as a causal diamond in the bulk (with the same time-like separated points but now the causal past and future of $p$ and $q$ respectively, include points also in the bulk, i.e, the diamond includes causal paths from $q$ to $p$ that extend in the bulk), any point in the wedge can be described only using the data on the time slice.  

AdS-Rindler wedge is considered as the dual to the boundary causal diamond of a finite spherical region. It is the bulk region that can be reconstructed from this boundary sub-region.\footnote{More precisely, what is dual to a boundary sub-region is the entanglement wedge but in the simple cases where the state is the vacuum, the entanglement wedge and the causal wedge coincide.} Since the algebra of operators in a sub-region of a quantum field theory in flat space-time is expected to be Type III$_{1}$\cite{Haag:1992hx}, one can conclude that the algebra of operators in the AdS-Rindler wedge is also Type III$_{1}$.
Without even referring to the boundary, one can see that the algebra should be Type III$_{1}$ since in the bulk one has a local quantum field theory on a curved space-time in the strict large $N$ limit. So a sub region like AdS-Rindler wedge will be Type III$_{1}$.

\subsection{Explicit construction of the algebra}

Once we have established the type of the algebra of operators in AdS-Rindler, $\mathcal{\tilde{A}}_{0}$, it follows that the Hilbert space that describes the system will have to be built out of a thermofield double state of the naive Hilbert space one would come up with. The reason for this can be understood if one thinks of the quantum field theory on AdS-Rindler space as a QFT at temperature $T = 1/2\pi$, which arises as a result of the Rindler horizon.

In the usual quantum field theory at zero temperature \cite{Streater:1989vi}, the way to construct a Hilbert space is by first choosing a vector that one is interested in, $\ket{\xi}$\footnote{It is taken to be the vacuum in most cases.}, and start applying with the modes in the algebra of the system. There will be infinite number of modes, but in most interesting systems it will be enough to act with finite but arbitrary number of modes.\footnote{Physically the intuition is that any experimenter will not be able to act with infinite number operators.} This will create for us the pre-Hilbert space. To create the Hilbert space, one requires any Cauchy sequence to converge, i.e, one adds the limit points of any Cauchy sequence, 
$$ \{\ket{\psi_{n}}\}_{n \in \mathbb{N}}, $$
in the pre Hilbert space, this is called its Hilbert space completion. 

This Hilbert space, $\mathcal{H}_{[\xi]}$, will not involve states that differ from the original state, $\ket{\xi}$, by the action of infinite number of modes. But with such states, one can create other Hilbert spaces in the same way (by acting with arbitrary but finite number operators and taking the Hilbert space completion). These Hilbert spaces are totally independent\footnote{The fact that they are independent will necessarily depend on the algebra of operators.} and can be treated as different super selection sectors\cite{10.2307/1969463}.
Naively one would think that the full Hilbert space of the system would be the space that includes all the Hilbert spaces whose construction was given just now. 
$$ \mathcal{H}^{'}=\bigoplus_{[\xi]} \mathcal{H}_{[\xi]}$$
This Hilbert space, called non-separable Hilbert space, has uncountably infinite dimension and is hard to deal with. But, since the super selection rules apply to the 'small Hilbert spaces', it would be enough to consider only these super selection sectors whose dimension is countably infinite.     

It is this construction that fails in the case of a QFT at non zero temperature. The reason is thermal fluctuations will take us outside the super selection sectors. Thus, it seems that one necessarily have to deal with the non-separable Hilbert space. A way around this was found by von Neumann \cite{10.2307/1968823} by using the thermofield double state at temperature $T$ to construct a separable Hilbert space but now the algebra of operators will not act irreducibly on the Hilbert space. Rather than choosing some vector to build the Hilbert space on, now one chooses the thermofield double state at a given temperature and different super selection sectors corresponds to starting with thermofield double states at different temperatures. Thus thinking the AdS-Rindler wedge in the strict large $N$ limit, $N=\infty$, as a quantum field theory on a curved space-time at some temperature $T$, one uses the thermofield double state at temperature $T= 1/2 \pi$ to construct the Hilbert space on which the Type III$_{1}$ algebra will act.

The algebra $\mathcal{\tilde{A}}_{0}$ contains bulk fluctuations acting in the AdS-Rindler wedge. Consider the following metric for the AdS-Rindler\cite{Almheiri:2014lwa},
\begin{equation}
    ds^{2}=-(r^{2}-1)dt^{2} + \frac{dr^{2}}{r^{2}-1} + r^{2}dH_{d-1}^{2}
\end{equation}
where $dH^{2}_{d-1} = d\chi^{2} + \chi^{2} d\Omega_{d-2}^{2}$ is the standard metric on the hyperbolic ball $H_{d-1}$.

A free massless scalar field on this metric is given by,
\begin{equation}
    \phi(t,r,\alpha) = \sum_{\lambda} \int_{0}^{\infty}\frac{d\omega d\lambda}{4\pi^{2}} (f_{\omega, \lambda}(t,r,\alpha) a_{\omega, \lambda} + f^{\star}_{\omega, \lambda}(t,r,\alpha) a^{\dagger}_{\omega, \lambda} )  
\end{equation}
where the sum is over degeneracy for $\lambda$ and $f(t,r,\alpha)$ solves the Klein-Gordon equation and is of the form,
$$ f_{\omega, \lambda}(t,r,\alpha) = e^{i\omega t} Y_{\lambda}(\alpha)\psi_{\omega,\lambda}(r)$$
and $Y_{\lambda}(\alpha)$ is eigenfunction of the Laplacian on $H_{d-1}$ and $\psi_{\omega,\lambda}(r)$ is given in terms of hypergeometric functions up to some prefactors that make sure that it has the expected asymptotic behaviour. The elements of $\mathcal{\tilde{A}}_{0}$, $F(a,a^{\dagger})$, are linear combinations of a finite string of operators like
$$(a^{\dagger}_{\omega_{1}, \lambda_{1}})^{i_{1}} ... (a^{\dagger}_{\omega_{r}, \lambda_{r}})^{i_{r}}(a_{\omega_{1}, \lambda_{1}})^{k_{1}} ... (a_{\omega_{s}, \lambda_{s}})^{k_{s}}. $$
 with complex coefficients.
 
Naively, one would start with the vacuum, a state annihilated by all the $a_{\omega, \lambda}$'s and then consider all states that are created by the action of the $a^{\dagger}_{\omega, \lambda}$'s.\footnote{From the full bulk perspective (in contrast to the just the AdS-Rinlder subregion), this vacuum is not well defined, it has a firewall (infinite stress energy tensor) at the entangling surface with the complement region. The reason is that there will be no entanglement between the two complementary regions and the two point functions for this state between the two regions will not have the correct short distance behaviour.} 
Considering the vacuum of the full bulk (the AdS vacuum), $\ket{vac}$, one can see that these modes are thermally populated, 
\begin{equation}
    \bra{vac}a^{\dagger}_{\omega, \lambda}a_{\omega^{'}, \lambda^{'}}\ket{vac}= \frac{1}{e^{2 \pi \omega}-1} \delta(\lambda-\lambda^{'})\delta(\omega-\omega^{'})
\end{equation}
But because of the thermal fluctuations, 
states that can only reached from the AdS-Rindler vacuum by the action of infinite number of $a^{\dagger}_{\omega, \lambda}$'s, should also be included. But, as remarked before, rather than considering the full non-separable Hilbert space, one will proceed to construct the Hilbert space on top of the thermofield double state. 

The AdS-Rindler vacuum is $ \ket{\Tilde{0}}=\bigotimes_{\omega,\lambda} \ket{\Tilde{0}}_{\omega\lambda}$
and an arbitrary state created by the action of arbitrary but finite number of $a^{\dagger}_{\omega, \lambda}$'s on each state $\ket{\Tilde{0}}_{\omega\lambda}$and take the tensor product over $\omega$ and $\lambda$,
$$    \ket{\tilde{j}}=\bigotimes_{\omega,\lambda} \ket{\tilde{j}_{\omega\lambda}}.$$
The Hilbert space $\mathcal{H}_{0}$ spanned by these states $\ket{\tilde{j}}$ is the non separable Hilbert space. We consider a state (here we assume that we discretized $\omega $ and $\lambda$ and take the continuum limit later when we take $n$ to infinity),
\begin{equation}\label{3.11}
    \ket{\tilde{\Psi}_{n}} = \frac{1}{\sqrt{\tilde{Z}_{n}}} \bigotimes_{\omega,\lambda}^{n} \sum_{j_{\omega\lambda}} e^{-\pi \omega j_{\omega\lambda}} \ket{\tilde{j}_{\omega\lambda}} \ket{\tilde{j}_{\omega\lambda}}^{'} 
\end{equation}
where $\tilde{Z}_{n}$ is such that  $\bra{\tilde{\Psi}_{n}}\ket{\tilde{\Psi}_{n}}=1$ and the state $\ket{\tilde{j}_{\omega\lambda}}^{'}$ is an element of $\mathcal{H}_{0}^{'}$, a copy of $\mathcal{H}_{0}$. Thus we define the thermofield double state $\ket{\tilde{\Psi}}$ as a map from $\mathcal{\tilde{A}}_{0}$ to $\mathbb{R}$ given by,
\begin{equation}\label{o}
\bra{\tilde{\Psi}}F(a,a^{\dagger})\ket{\tilde{\Psi}}= \lim_{n \rightarrow \infty} \bra{\tilde{\Psi}_{n}}F(a,a^{\dagger})\ket{\tilde{\Psi}_{n}}.
\end{equation}

\footnote{There should also be an upper limit on the sum over $j_{\omega, \lambda}$ which we did not emphasized here. A more careful statement would be to consider $\ket{\tilde{\Psi}_{n}} = \frac{1}{\sqrt{\tilde{Z}_{n}}} \bigotimes_{\omega,\lambda}^{n} \ket{\omega,\lambda}$ and define $\ket{\omega,\lambda}_{k}= \sum_{j_{\omega\lambda}}^{k} e^{-\pi \omega j_{\omega\lambda}} \ket{\tilde{j}_{\omega\lambda}} \ket{\tilde{j}_{\omega\lambda}}^{'} $ and  $\ket{\omega,\lambda}$ is defined by $\bra{\omega\lambda}F(a_{\omega\lambda},a_{\omega\lambda}^{\dagger})\ket{\omega\lambda}= \lim_{k \rightarrow \infty} \bra{\omega\lambda}_{k}F(a_{\omega\lambda},a_{\omega\lambda}^{\dagger})\ket{\omega\lambda}_{k}.$}
Measurements made by an observer restricted to this
bulk sub region is thus given by such expectation value.
The Hilbert space $\mathcal{H}_\Psi \subset \mathcal{H}_{0}$ is generated by states given by the action of a finite but arbitrary string of $a$ and $a^{\dagger}$’s on $\ket{\tilde{\Psi}}$
and their Hilbert space completion.  

The boundary analog of the algebra of operators and Hilbert space construction can be discussed following the extrapolate dictionary\cite{Banks:1998dd,Harlow:2011ke},
$$ \lim_{r \rightarrow \infty} r^{\Delta}\phi(r,x) = O(x),$$
$\Delta$ being the conformal dimension of the conformal field theory single trace operator $O$. Then it follows that,
\begin{equation}
    O_{\omega\lambda} = \int dtd\alpha\; e^{-i\omega t}Y^{\star}_{\lambda}(\alpha) O(t,\alpha) = N_{\omega\lambda}a_{\omega\lambda}
\end{equation}
Then, the non-separable Hilbert space $\mathcal{H}_{0}$ is spanned by $\ket{j}=\bigotimes_{\omega\lambda}\ket{j_{\omega\lambda}}$ constructed from the vacuum of the boundary subregion, $\ket{0}=\bigotimes_{\omega\lambda}\ket{0}_{\omega\lambda}$, by the action of $j_{\omega\lambda}$ raising operators, $O^{\dagger}_{\omega\lambda}.$

We also define a thermofield double state $\ket{\Psi}$ as a map form $\mathcal{B}_{0}$, the algebra for single trace operators, to $\mathbb{R}$ exactly like \ref{3.11} and \ref{o} but now $a_{\omega\lambda}$ and $a^{\dagger}_{\omega\lambda}$ are replaced by $O_{\omega\lambda}$ and $O^{\dagger}_{\omega\lambda}$.

Single trace operators are gauge invariant operators that can be constructed by considering operators written in terms of trace of the fields and their derivatives with no explicit $N$ dependence in matrix theories. In these theories, where the fields transform in the adjoint representation of, for example $SU(N)$ or $U(N)$, the fields can be normalized such that the action is given with an explicit $N$ factor multiplied by a gauge invariant term with no explicit $N$ dependence. Well known examples include free Yang Mills,  $\mathcal{N}=4$ super Yang Mills and so on.
These operators have $O(N)$ one point function and two point functions with a leading $O(N^{2})$ disconnected term. Thus to make sense of the large $N$ limit, one needs to consider the subtracted single trace operators, $\tilde{O}=O-\langle O \rangle$. These operators close and are generalized free fields in the large $N$ limit. Thus, since we are studying the $N= \infty$ theory, one should use $\Tilde{O}$ to construct $\mathcal{B}_{0}$ which are operators with a well defined large $N$ limit.

The Hilbert space, $\mathcal{H}_{\Psi}$ is now constructed from the action of operators in $\mathcal{B}_{0}$ on $\ket{\Psi}$ and limit points of a Cauchy sequence of states.

The algebra $\mathcal{B}_{0}$ should also be completed in the weak sense, i.e, if a sequence of expectation values converges, $\lim_{n \rightarrow \infty} \bra{\Psi}F_{n}\ket{\Psi}=\bra{\Psi}F\ket{\Psi}$ for $F_{n} \in \mathcal{B}_{0}$, then one adds the operator $F$ to get the von Neumann algebra $\mathcal{B}$. This is the algebra of single trace operators.

But this algebra of single trace operators is not all there is, there is a conserved charge associated with the conformal Killing vector, $\chi^{\mu}$, that preserve the causal diamond of the spherical boundary sub region and its complement. It is given by an integral that involves the stress energy tensor, $\chi^{\mu}$ and the normal vector to the boundary Cauchy surface. Since it is a symmetry it annihilates the full vacuum, $\ket{vac}$. In the bulk this is the boost operator that preserves Rindler horizon. The integral can be naturally divided into a part that acts in the subregion and a part that acts in the complement, $H_{\chi}= 2 \pi(K-\Bar{K})$. The operator $K$ is the boost operator of the boundary subregion and the algebra of single trace operators will not capture it. In the strict large $N$ limit, this corresponds in the bulk to the area of the horizon, loosely speaking\footnote{There are several subtleties concerning the diverging behaviour of the area term which we will discuss in the following sections.}, which is also not captured by the bulk fluctuations.

In the normalization that single trace operator, $O$, have $O(N)$ one point function and leading disconnected $O(N^{2})$ two point function. On the other hand, the stress energy tensor will have the same form as the action, a gauge invariant term multiplied by $N$. The generator of the conformal transformation that leaves the diamond of the boundary subregion $B$ invariant is given by,
\begin{equation}
    K = \int_{B}T_{\mu \nu}\chi^{\mu}dB^{\nu}
\end{equation}
and will also have the explicit $N$ dependence. This explicit $N$ dependence will give $K$ one point and connected two point functions of $O(N^{2})$. Thus these quantities will become divergent in the large $N$ limit. In addition, there is also a UV divergence arising from the UV degrees of freedom close to the boundary of the subregion of the boundary CFT. We will discuss this UV divergence later in this section and the next section. A simple subtraction like what we did for the single trace operators will not result in a generalized free field, since the two point function will still be $O(N^2)$. Thus one defines the operator,
\begin{equation}\label{3.9}
   \tilde{X} = \frac{K-\langle K\rangle}{N}
\end{equation}
Up till now one can draw a direct parallel between the case of large black holes in AdS, with the Hamiltonian being the corresponding $K$\cite{Witten:2021unn}. The Hamiltonian is also given as an integral of the stress energy tensor and has $O(N^{2})$ one point and connected two point function (above Hawking-Page temperature)\footnote{One argument for the Type III$_{1}$ nature of the AdS-Rindler wedge and also black hole exterior in AdS is the continuous spectrum of the modular Hamiltonian (boost operator) (the Hamiltonian for the black hole case). This continuous spectrum of the Hamiltonian is in fact associated to the appearance of horizon in the bulk\cite{Festuccia:2006sa}.}. The only difference would be in the case of the black hole, there is a finite horizon area. But in the present case, the horizon is of infinite volume. Thus, even though for the Hamiltonian one would be satisfied with \ref{3.9}, for the present case there will still be an additional divergence coming from horizon which was not present for the black hole and strictly speaking, the mode $\tilde{X}$ is not yet well defined.  

From the bulk perspective, this can be seen from the computation of $\langle K\rangle$ and $\langle K^{2}\rangle-\langle K\rangle^{2}$ in \cite{Verlinde:2019ade}. To leading order in $G_{N}$,
\begin{equation}\label{3.10}
  \langle K\rangle = \langle(K-\langle K\rangle)^{2}\rangle = \frac{A(\Sigma)}{4G_{N}}
\end{equation}
where $A(\Sigma)$ is the volume of the Rindler horizon $\Sigma$ which is a $d-1$ dimensional hyperbolic hyper-surface. Following the normalization \ref{3.9} and using $G_{N} \sim 1/N^{2}$ one gets $\langle \tilde{X}\rangle=0$ and $\langle \tilde{X}^{2}\rangle \sim A(\Sigma)$. The two point function is infinite since $A(\Sigma)$ is infinite. This is as a result of the boundary UV divergence that was mentioned earlier.

Even though $\tilde{X}$ is well defined $\tilde{X}^{2}$ is not well defined, it has divergent expectation value. If there was any way to renormalize it and define a new operator $X$ with a well defined large $N$ limit for the one and two point function\footnote{Here we assume large $N$ factorization for $X$, so it is enough to renormalize the two point function.}, then one can include to our strict large $N$ limit algebra of operators, the bounded functions of this mode and get the full algebra $\mathcal{A}=\mathcal{\tilde{A}}_{0}\otimes \mathcal{A}_{X}$. This algebra of operators will act on an extended Hilbert space, $\mathcal{H}=\mathcal{H}_{\Psi}\otimes \mathcal{L}^{2}(\mathbb{R})$ which in general include some entangled state between square intergrable functions of $X$ and states in $\mathcal{H}_{\Psi}$.

\subsection{The renormalization of the infinite volume }
We now claim that there is a natural renormalization of this additional divergence mentioned in the previous section. We start by noticing \ref{3.10} that the expectation value and the fluctuations of $K$ are given by the area of a $d-1$ dimensional hyperbolic surface with boundary. A hyperbolic manifold with boundary will have, in general, infinite volume. But there is a canonical renormalization of this volume which would follow from looking at the Einstein-Hilbert action of these manifolds. 

Einstein equations imply that the bulk part of the Einstein-Hilbert action is given by the volume of the manifold (up to some prefactors)
. For a non compact manifold $M$, this action can be shown to be infinite. But one can consider a finite sub manifold $N \subset M$ and take the action on this sub manifold. In the limit $\partial N$ goes to $\partial M$, which is at infinity, one can see that the action diverges in terms of local quantities on $\partial N$ which are invariant functions of the induced metric, more specifically the first and second fundamental forms. Thus adding counter terms to this action and renormalizing it will enable one to associate a canonical finite volume to the manifold.\footnote{In 3 dimensions it is canonical up to the Weyl anomaly of two dimensional CFT. This anomaly is absent in even dimensions.} In particular, this would imply to a renormalization where one subtracts the divergent terms and then take the limit $\partial N$ going to $\partial M$. Thus giving a notion of renormalized volume for the hyperbolic manifold $M$.

The action for a manifold $M$ with Euclidean signature is given by,
\begin{equation}
    S=-\frac{1}{16\pi G_{N}}\int_{M}d^{d-1}x \sqrt{g}(R - 2\Lambda)-\frac{1}{8\pi G_{N}}\int_{\partial M}d^{d-2}x\sqrt{\gamma}K_{\gamma}
\end{equation}
where $\Lambda$ is the cosmological constant that is assumed to be negative, $g$ and $R$ are the metric and Ricci scalar on $M$ while $\gamma$ and $K_{\gamma}$ are the induced metric and the mean curvature on its boundary.  



The renormalization procedure is discussed in several papers\cite{Henningson:1998gx,Krasnov:2006jb} but we will revise it here. It essentially involves four steps. The first is to take $M$ to be in the metric,
\begin{equation}
    ds^{2}=\frac{1}{4\rho^{2}}d\rho^{2}+\frac{h_{ij}}{\rho}dx_{i}dx_{j}
\end{equation}
as $\rho$ goes to zero (the boundary), $h$ goes to $\gamma$, the boundary metric. 
Then one solves Einstein's equation for the above metric.
\begin{equation}
  \begin{split}
    \rho(2h^{''}-2h^{'}h^{-1}h^{'}+Tr(h^{-1}h^{'})h^{'})
        + R_{\gamma}-(d-2)h^{'}-Tr(h^{-1}h^{'})h&=0\\
       (h^{-1})^{jk}(D_{i}h^{'}_{jk}-D_{k}h^{'}_{ij})&=0\\
       Tr(h^{-1}h^{''})-\frac{1}{2}Tr(h^{-1}h^{'}h^{-1}h^{'})&=0
   \end{split}
\end{equation}
where $R_{\gamma}$ is the boundary Ricci scalar and $D_{i}$ is covariant derivative in the boundary metric while prime is derivative with respect to $\rho$. We have also used $\Lambda = \frac{-d(d-1)}{2}$. 

The important next step is to note that
the above 
equations can be solved by considering the following expansion for $h$,
\begin{equation}\label{3.14}
    h=\gamma + \rho \gamma_{2} +\rho^{2}\gamma_{4}+... 
\end{equation}
 where for even $d$, there is an additional term at the $\rho^{d/2}$ order which is $\rho^{d/2}$log$\rho$ $ \kappa_{2}$. $\gamma_{k}$ is given covariantly in terms of $\gamma$ and contains $k$ derivatives with respect to $x_{i}$.
 
 Considering the simplest case of hyperbolic 3 manifolds, i.e, taking the full bulk to be $AdS_{5}$, Einstein's equations become $R-2\Lambda=-4$ and one can see that the action is just the volume of the manifold $M$\footnote{There are also additional boundary terms but we are going to use local boundary counter terms to renormalize the action and, for 3 manifolds, these boundary terms are cancelled exactly by the counter terms.}. Now we consider the action of a sub manifold $N_{\epsilon} \subset M$ that is regularized at  $\rho = \epsilon$ for some cut off $\epsilon>0$. The action of $N_{\epsilon}$ will have both bulk and boundary term, at $\rho=\epsilon$. The expansion \ref{3.14} implies a similar expansion for $\sqrt{g}$ and the action will be as follows,
 \begin{equation}
     S=\frac{V(N_{\epsilon})}{4\pi G}= \frac{1}{4\pi G} \int_{\partial N_{\epsilon}} d^{2}x\; \frac{1}{\epsilon}a_{0}+ \text{log}\epsilon \;a_{2} + L_{finite}.
 \end{equation}
 where $a_{0}=\sqrt{\gamma}$ and $a_{2}=-\frac{1}{4} \sqrt{\gamma}R_{\gamma}$. Thus adding counter terms to remove the divergent terms and taking the limit $\epsilon \rightarrow 0$, one gets a renormalized volume of the manifold $M$.
 
 For our present case, this hyperbolic manifold is the horizon of the AdS-Rindler subregion. If we renormalize $\tilde{X}^{2}$ by subtracting the above divergent terms, then the renormalized mode $X$ is well defined with finite one and two point functions and the algebra given above, $\mathcal{A}=\mathcal{\tilde{A}}_{0}\otimes \mathcal{A}_{X}$ makes sense. In addition, this provides a canonical renormalization of the Ryu-Takayanagi surface for a spherical region in the boundary. Since the renormalization of the volume of hyperbolic manifolds is applicable for any dimensions \cite{Balasubramanian:1999re,Krasnov:2006jb}, the renomalization scheme can be used for boundary CFTs in other dimensions.

\section{Adding gravity and making the algebra Type II$_{\infty}$}

This section closely follows the discussion in \cite{Witten:2021unn} for the eternal black hole in AdS. To move away from the strict large $N$ limit, we will have to include in $\mathcal{\tilde{A}}_{0}$ linear combinations of single trace operators with coefficients that have asymptotic power expansion in $1/N$ around $1/N = 0$. In addition, one have to add $1/N$ corrections to $X$. The way to do this is to note that modular Hamiltonian at finite $N$ generates modular time translation within the causal diamond of the boundary sub region.
\begin{equation}\label{4.1}
    [\frac{K-\langle K\rangle}{N},O]=\frac{1}{N}\partial_{\tau}O.
\end{equation}

Thus one adds to $X$ an operator that has well defined large $N$ limit and also has no divergence coming from infinite horizon area, i.e, $H_{\chi}$. Checking with \ref{4.1}, one has a new operator, $X+\frac{H_{\chi}}{2 \pi N}$ that generates time translation for the single trace operators. 
Any other operator that one can add consistent with \ref{4.1}, like operators from the complement region, can be removed by making use of the conjugate operator of $X$, that is $\Pi=\frac{d}{idX}$. By conjugating the algebra with $e^{i\Pi \hat{O}/N}$, one can remove  an additional operator from the complement, like $\hat{O}/N$\cite{Witten:2021unn}.\footnote{The reason is that these algebras are defined up to conjugation.} Thus, even at perturbative order in $1/N$, the new mode does not have divergences and is well defined.

The bounded functions for this mode will form the algebra $\mathcal{A}_{X+\frac{H_{\chi}}{2 \pi N}}$. Since this enlargement of the algebra upgrades the center mode to an outer automorphism for $\mathcal{\tilde{A}}_{0}$, as is well known in the axiomatic quantum field theory literature and very well elaborated in a paper, the full algebra will be a Type II$_{\infty}$ crossed product algebra $\mathcal{\tilde{A}}_{0}\ltimes \mathcal{A}_{X+\frac{H_{\chi}}{2 \pi N}}$. This algebra again will act on the Hilbert space, $\mathcal{H}=\mathcal{H}_{\Psi}\otimes \mathcal{L}^{2}(\mathbb{R})$ but now one can associate a density matrix for any state. A certain class of  states one can consider are of the form $\ket{\tilde{\Psi}}=\ket{\Psi}\otimes g(X)^{1/2}$. Using such states, a trace can be defined for what are called trace class operators\cite{Witten:2021unn}. An element of the crossed product algebra, $\tilde{O}=\int ds\; O(s)e^{isY}$ where $Y=X+\frac{H_{\chi}}{2 \pi N}$ is trace class if the trace of the operator defined as,
$$    Tr(\tilde{O}) = \bra{\tilde{\Psi}}\tilde{O} \tilde{K}^{-1}\ket{\tilde{\Psi}}, $$
is finite, where $\tilde{K}=\frac{g(Y)}{e^{Y}}$. This implies the density matrix for such states is in fact $\tilde{K}$. One can also define a density matrix, $\rho_{\Phi}$, for a general state $\ket{\tilde{\Phi}} \in \mathcal{H}$. 
Thus, it is possible to define a von Neumann entropy, $\mathcal{S}_{\Phi}
=-Tr\rho_{\Phi}\text{log} \rho_{\Phi}=-\bra{\tilde{\Phi}}
\text{log} \rho_{\Phi} \ket{\tilde{\Phi}}$, can be associated to
any state in the Hilbert space.

\section{Discussion}

This analysis has been for AdS-Rinlder wedge at perturbative order in $1/N$ around $N=\infty$. But for finite $N$ , the algebra of the AdS-Rindler is expected
to be again Type III$_{1}$. The reason is from the boundary
perspective, one is looking at the algebra of a local region of a quantum field theory which is in general Type
III$_{1}$. On the other hand, the space-time description of
the theory will not be precise for finite value say $N=243$. Thus the renormalization procedure for the area of the Rindler horizon will make less
and less sense as one backs away from the N going to
infinity limit. Thus we expect that, to make sense of the
theory non perturbatively, one would have to work with
 $\Tilde{X}$ and not the renormalized mode; this restores the Type III$_{1}$ nature of the boundary sub region for finite $N$. But still a more careful analysis has to be done to understand this better.

It would also be interesting to compare the von Neumann entropy given in the previous section to the usual entropy proposed by Hubeny, Rangamani and Takayanagi\cite{Hubeny:2007xt}.

\acknowledgments
I would like to thank Kyriakos Papadodimas for the useful discussions and a careful reading of the article. I also want to thank the CERN-TH for their hospitality during the preparation of this paper. The research is partially supported by INFN Iniziativa Specifica String Theory and Fundamental Interactions.

\bibliographystyle{JHEP}
\bibliography{4}

\end{document}